\documentclass[aps,prb,showpacs,twocolumn]{revtex4}
\usepackage{txfonts}
\usepackage{graphicx}

\begin{document}

\title{Weak coupling $d$-wave BCS superconductivity and unpaired electrons in overdoped La$_{2-x}$Sr$_x$CuO$_4$ single crystals}

\author{Yue Wang, Jing Yan, Lei Shan, and Hai-Hu Wen}\email{hhwen@aphy.iphy.ac.cn }

\affiliation{National Laboratory for Superconductivity, Institute of
Physics and Beijing National Laboratory for Condensed Matter
Physics, Chinese Academy of Sciences, P.O. Box 603, Beijing 100080,
People's Republic of China}

\author{Yoichi Tanabe, Tadashi Adachi, and Yoji Koike}

\affiliation{Department of Applied Physics, Graduate School of
Engineering, Tohoku University, 6-6-05 Aoba, Aramaki, Aoba-ku,
Sendai 980-8579, Japan}

\begin{abstract}
The low-temperature specific heat (SH) of overdoped
La$_{2-x}$Sr$_x$CuO$_4$ single crystals ($0.178\leqslant
x\leqslant0.290$) has been measured. For the superconducting samples
($0.178\leqslant x\leqslant0.238$), the derived gap values (without
any adjusting parameters) approach closely onto the theoretical
prediction $\Delta_0=2.14k_BT_c$ for the weak-coupling $d$-wave BCS
superconductivity. In addition, the residual term $\gamma(0)$ of SH
at $H=0$ increases with $x$ dramatically when beyond $x\sim0.22$,
and finally evolves into the value of a complete normal metallic
state at higher doping levels, indicating growing amount of unpaired
electrons. We argue that this large $\gamma(0)$ cannot be simply
attributed to the pair breaking induced by the impurity scattering,
instead the phase separation is possible.
\end{abstract}
\pacs{74.25.Bt,74.20.Rp, 74.25.Dw, 74.72.Dn}

\maketitle

\section{INTRODUCTION}

For hole-doped cuprates, it is now generally perceived that the
superconducting state has robust $d$-wave symmetry.\cite{CCT} In the
underdoped region, due to the presence of the pseudogap and other
possible competing orders,\cite{JO,DAB} the measured quasipartcle
gap may not reflect the real superconducting gap. In contrast, in
the overdoped region, the normal state properties can be described
reasonably well by the Fermi liquid picture,\cite{CP} although still
with electronic correlation to some extent.\cite{SN} Under this
circumstance, one may think that the overdoped cuprate provides a
clean gateway to the intrinsic high-$T_c$ superconducting state. To
accumulate experimentally accessible parameters, such as the
superconducting gap, and compare them with the mean field BCS
prediction in this very region is thus expected to be particularly
valuable.

Another puzzling point in the overdoped cuprates is that the
superfluid density $\rho_s$ determined by muon spin relaxation
($\mu$SR) technique decreases just as the transition temperature
$T_c$ when beyond a critical doping point
$p_c\sim0.19$.\cite{YJU,CN,CB} This is actually not demanded by the
BCS theory. The decrease of $\rho_s$, first reported in
Tl$_2$Ba$_2$CuO$_{6+\delta}$ (Tl2201) and subsequently confirmed in
other families of cuprates,\cite{CB,JPL} was attributed to the
unpaired carriers at $T\rightarrow0$ in overdoped
cuprates.\cite{YJU} Similar conclusion was also drawn from studies
of the optical conductivity\cite{JS,SU} and magnetization.\cite{HHW}
Recently, the Meissner volume fraction was revealed to decrease as
$T_c$ with increasing doping in overdoped La$_{2-x}$Sr$_x$CuO$_4$
(LSCO) and the result was suggested to be consistent with the
occurrence of a phase separation into superconducting and
normal-state regions.\cite{YT} It is thus highly desired to use the
specific heat (SH) which is very sensitive to the quasiparticle
density of states (DOS) at the Fermi level to directly probe these
unpaired charge carriers.

In this paper we shall address these two issues by the
low-temperature SH which has established its importance to identify
the pairing symmetry in high-$T_c$ cuprates over the past
decade.\cite{NEH} Recently, the quantitative analysis shows that it
provides a bulk way to obtain the magnitude of the superconducting
gap.\cite{YW,HHW3} By analyzing the field-induced SH, it is found
that the pairing symmetry in the overdoped regime (up to $x=0.238$)
is still $d$-wave and the derived gap values $\Delta_0$ approach
closely onto the theoretical prediction of the weak-coupling
$d$-wave BCS superconductivity. Our data also reveal a quick growing
of the residual term $\gamma(0)$ of SH at $T\rightarrow0$ with
increasing doping, which cannot be simply attributed to the pair
breaking induced by the impurity scattering.

\section{EXPERIMENT}

\begin{figure}
\includegraphics[width=6cm]{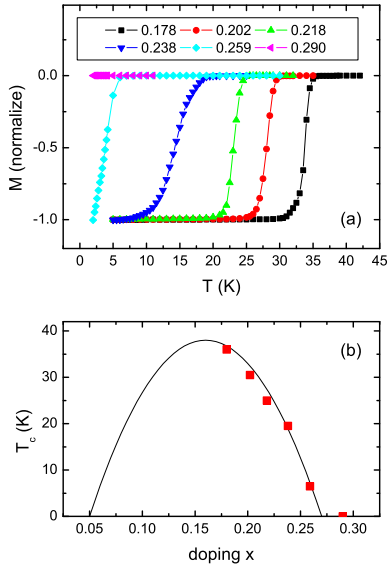}
\caption{(color online) (a): Temperature dependence of DC
magnetization measured in ZFC mode at 10 Oe. The curves are
normalized to unity with the value at the lowest temperatures. (b):
Doping dependence of $T_c$ (squares). The empirical formula (see
text) is plotted as the solid line. } \label{fig1}
\end{figure}

Single crystals of LSCO were grown by the traveling-solvent
floating-zone method. Details of the sample preparation have been
given elsewhere.\cite{YT,TA} The Sr content of the sample, $x$,
taken as the hole concentration $p$, was determined from the
inductively coupled plasma measurement. Figure 1(a) shows the dc
magnetization curves measured in 10 Oe after the zero-field cooling
(ZFC) mode using a SQUID magnetometer, where the onset of the
diamagnetic signal was defined as the critical temperature $T_c$.
Six samples have been measured, with $x$ = 0.178, 0.202, 0.218,
0.238, 0.259, 0.290 and $T_c$ = 36, 30.5, 25, 19.5, 6.5, and below
1.7 K, respectively. As shown in Fig. 1(b), the $T_c$ can be
described well by the empirical formula\cite{JLT}
$T_c/T_c^\mathrm{max}=1-82.6(x-0.16)^2$ with $T_c^\mathrm{max}=38$
K. The SH measurements were performed on an Oxford Maglab cryogenic
system using the thermal relaxation technique, as described in
detail previously.\cite{HHW2} The temperature was down to 1.9 K and
the magnetic field was applied parallel to $c$-axis in the
measurements.

\section{RESULTS AND DISCUSSION}

\begin{figure*}
\includegraphics[width=19cm]{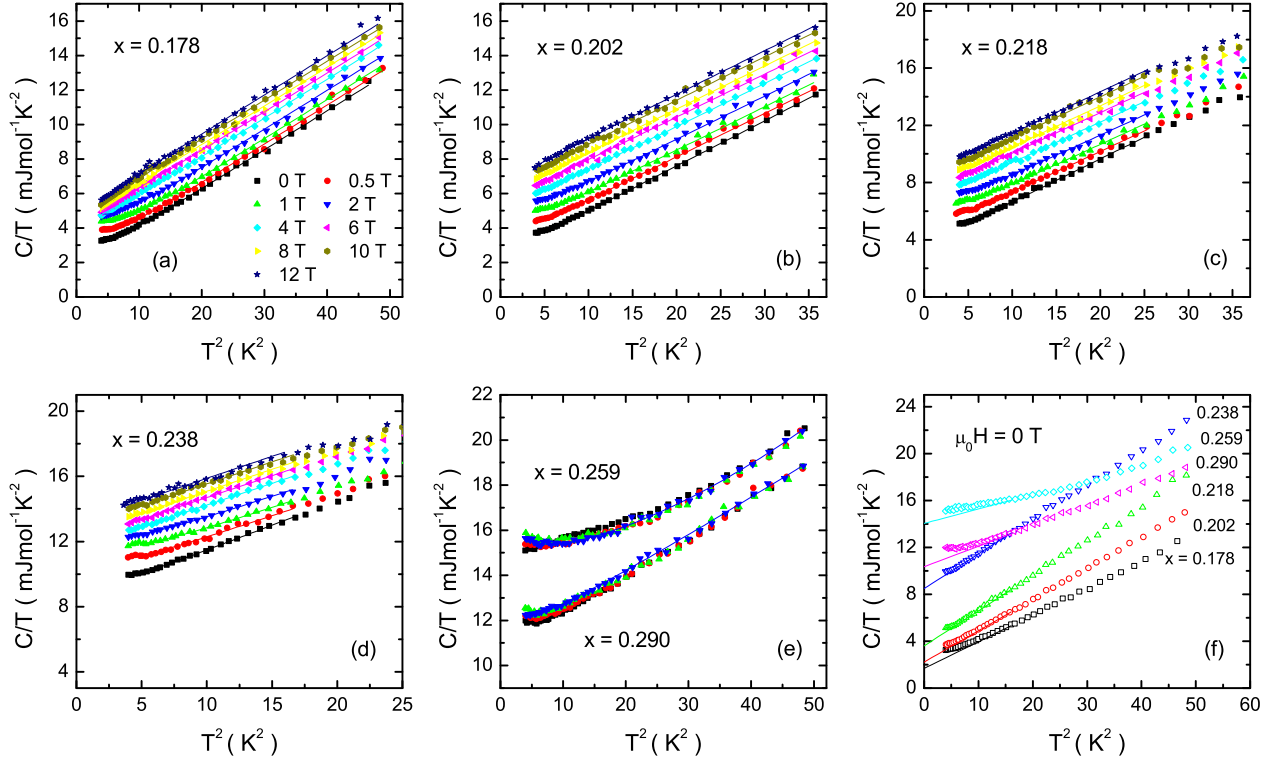}
\caption {(color online) Temperature and magnetic field dependence
of SH in $C/T$ vs $T^2$ plot. (a)-(e): Raw data for all six samples
(symbols). $\mu_0H$ varies up to 12 T for $0.178\leqslant
x\leqslant0.238$ while up to 2 T for $x=0.259$ and 0.290. The solid
lines represent the the theoretical fit (see text). The fits are
limited to $T=7$, 6, 5, and 4 K for $x=0.178$, 0.202, 0.218, and
0.238, respectively. For $x=0.259$ and 0.290, the fits are ranged to
7 K. (f): Replot the data at $\mu_0H=0$ T for all samples (symbol:
$\square=0.178$, $\medcirc=0.202$, $\vartriangle=0.218$,
$\triangledown=0.238$, $\lozenge=0.259$, $\vartriangleleft=0.290$).
The dotted lines are extrapolations of the data down to $T=0$ K with
the Schottky anomaly subtracted.} \label{fig2}
\end{figure*}

The raw data of SH for all six samples in various $H$ at $T<7$ K are
shown in Fig. 2. To separate the electronic SH from other
contributions, the data are fit to $C(T, H)=\gamma
T+C_\mathrm{ph}T+C_\mathrm{Sch}(T, H)$, where $C_\mathrm{ph}T=\beta
T^3$ is the phonon SH. $C_\mathrm{Sch}(T, H)$ is the two-level
Schottky anomaly with the form $D/T^2$ in $H=0$ and
$fx^2e^x/(1+e^x)^2$ ($x=g\mu_BH/k_BT$) in $H\neq0$, where $\mu_B$ is
the Boher magneton, $g$ the Land\'{e} factor, and $f$ the
concentration of spin-1/2 particles. The first linear-$T$ term,
$\gamma T$, contains the electronic SH and resides in the heart of
the present study. As demonstrated by the solid lines in Fig.
2(a)-(e), all data sets can be described reasonably well by the
above expression. For $C_\mathrm{ph}$ and the Debye temperature
$\Theta_{D}$ the values derived here are in consistent with previous
reports on the sample at similar doping level. \cite{SJC,SN} For
$C_\mathrm{Sch}$, the yielded $f$ is relatively constant at high
fields with an averaged value $\sim3$ mJmol$^{-1}$K$^{-1}$ for
different samples. This low $f$ reflects the small contribution of
$C_\mathrm{Sch}$ to the total SH and assures the reliable
determination of $\gamma$.

\begin{figure}
\includegraphics[width=8cm]{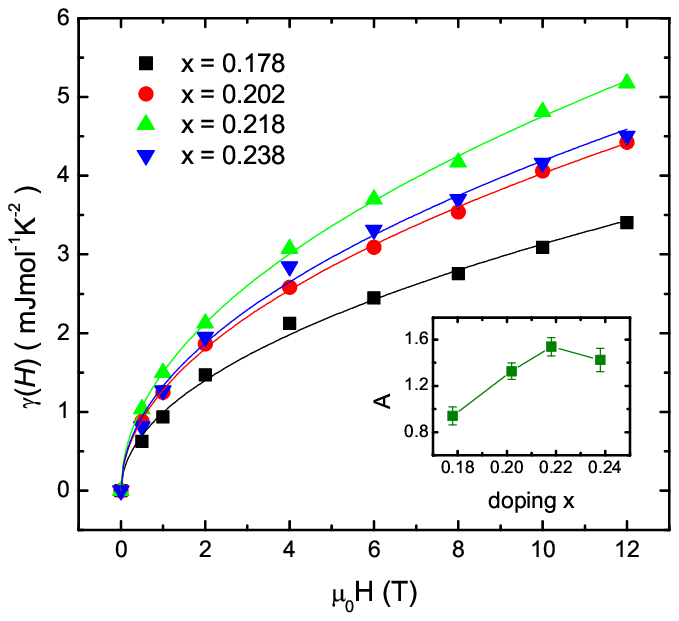}
\caption{(color online) Coefficient of the field-induced linear-$T$
specific heat for $0.178\leqslant x\leqslant0.238$,
$\gamma(H)=\gamma-\gamma(0)$ at $T=0$ K (symbols). The solid lines
are the fits to $\gamma(H)=A\sqrt{H}$. Inset: Doping dependence of
the prefactor $A$ (mJmol$^{-1}$K$^{-2}$T$^{-0.5}$).} \label{fig3}
\end{figure}

In zero-field, after removing the Schottky term $C_\mathrm{Sch}$ and
by doing a linear extrapolation to the data shown in Fig. 2(f) to
$T=0$ K we can determine the residual term $\gamma(0)$ of SH. By
increasing $H$, an increase in $\gamma$ is observed for
$0.178\leqslant x\leqslant0.238$, as shown in Fig. 2(a)-(d),
corresponding to $\gamma=\gamma(0)+\gamma(H)$ with $\gamma(H)$ the
coefficient of the field-induced SH. For $d$-wave superconductors,
it was theoretically pointed out that the $\gamma(H)$ is
proportional to $\sqrt{H}$ due to line nodes of the gap,\cite{GEV}
which has been confirmed in several experiments.\cite{NEH,HHW2}
Figure 3 summarizes the field dependence of the $\gamma(H)$ for the
overdoped LSCO. It is clear that for all four samples, the
$\gamma(H)$ is well described by $A\sqrt{H}$ with $A$ a
doping-dependent constant, as exemplified by the solid lines in Fig.
3. This indicates that in overdoped LSCO the gap remains robust
$d$-wave symmetry.

\begin{figure}
\includegraphics[width=7cm]{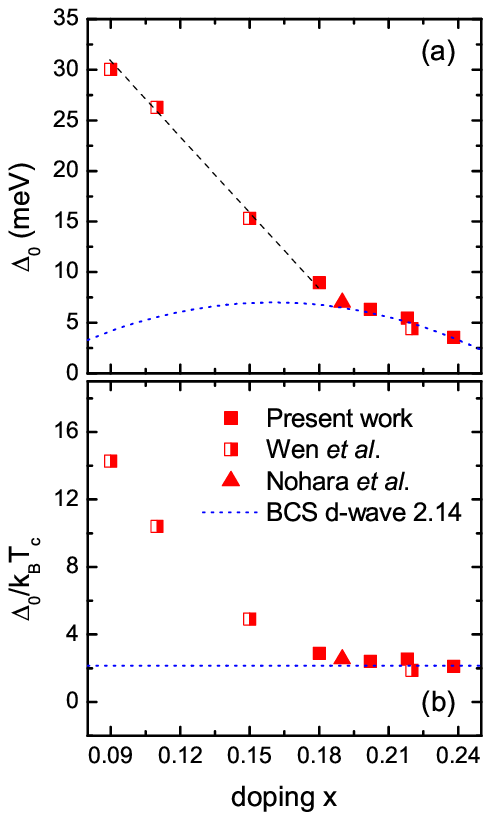}
\caption {(color online) Doping dependence of the superconducting
gap $\Delta_0$ obtained from SH measurements. (a): $\Delta_0$ vs
$x$. The dashed line is a guide to the eye. (b): $\Delta_0/k_BT_c$
vs $x$. The values from Ref. 16 (half-filled squares) and Ref. 25
(triangle) are also included. The weak-coupling d-wave BCS value,
$\Delta_0=2.14k_BT_c$ ($\Delta_0/k_BT_c=2.14$) is plotted as a
dotted curve (horizontal line) in (a) ((b)). For $x\geqslant0.19$,
the experiments are consistent with the BCS prediction.}
\label{fig4}
\end{figure}

Next one can further obtain the gap magnitude by investigating
$\gamma(H)$ quantitatively. Fundamentally, $\gamma(H)$ arises from
the Doppler shift of the quasiparticle spectrum near the nodes due
to the supercurrent flowing around vortices and thus directly
relates to the slope of the gap at the node,
$v_\Delta=2\Delta_0/\hbar k_F$ with $\Delta_0$ the $d$-wave maximum
gap in the gap function $\Delta=\Delta_0\cos(2\phi)$, $k_F$ the
Fermi vector near nodes (taking $\sim 0.7$ $\AA^{-1}$ as obtained
from ARPES\cite{TY}). Explicitly, the relation between $v_\Delta$
and the prefactor $A$ is given by
\begin{equation}
A=\frac{4k_{B}^{2}}{3\hbar}\sqrt{\frac{\pi}{\Phi_{0}}}\frac{nV_{mol}}{d}\frac{a}{v_{\Delta}}\label{eq:1}
\end{equation}
where $\Phi_0$ is the flux quantum, $n$ the number of CuO$_2$ planes
per unit cell, $d$ the $c$-axis lattice constant, $V_\mathrm{mol}$
the volume per mole, and $a=0.465$ for a triangular vortex
lattice.\cite{MC,IV} Inset of Fig. 3 shows the doping dependence of
$A$ by fitting the data to $\gamma(H)=A\sqrt{H}$. Thus, with the
known parameters for LSCO ($n=2$, $d=13.28$ $\AA$,
$V_\mathrm{mol}=58$ cm$^{3}$), the doping dependence of $v_{\Delta}$
and $\Delta_{0}$ can be derived without any adjusting parameters
according to Eq. (1). In this way we extracted the gap values
$\Delta_0=9.2\pm0.7$, $6.6\pm0.3$ and $5.6\pm0.3$ meV for $x=0.178$,
0.202 and 0.218, respectively (for $x=0.238$, the observed $A$
should be corrected due to the volume correction which will be
addressed later). It can be seen immediately that $\Delta_0$
decreases with increasing doping in the overdoped LSCO, concomitant
with the decrease of $T_c$. The same trend of $T_c$ and $\Delta_0$
with overdoping implies that the suppression of superconductivity
mainly comes from the decrease in the pairing gap. Figure 4 plots
the doping dependence of $\Delta_0$, together with the value
extracted from SH in underdoped and optimal doped LSCO single
crystals.\cite{HHW3,MN} For comparison, the weak-coupling $d$-wave
BCS gap relation $\Delta_0=2.14k_BT_c$ is also plotted as a dotted
curve in Fig. 4(a) and a dotted horizontal line in Fig.
4(b),\cite{HW} where $T_c$ is determined by the empirical formula
described before. Remarkably, beyond $x\sim0.19$, the experimental
data approaches closely onto the theoretical prediction, revealing a
strong evidence for the weak coupling $d$-wave BCS
superconductivity. Previous results about $\Delta_0$ determined by
scanning tunnelling spectroscopy\cite{TN} and penetration depth
measurements\cite{CPTX} are in excellent quantitative agreement with
our present results, which strongly supports the validity of the
present analysis.

\begin{figure}
\includegraphics[width=7cm]{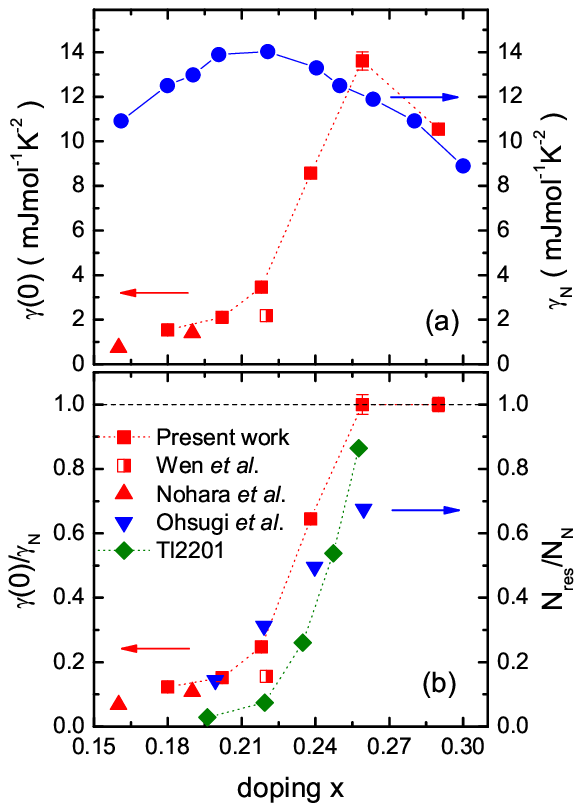}
\caption {(color online) (a): Doping dependence of the $\gamma(0)$
in zero field (filled squares) and the normal state SH coefficient,
$\gamma_N$ (circles). The $\gamma(0)$ from Ref. 19 (half-filled
square) and Ref. 25 (up-triangles) are also shown. The $\gamma_N$ is
quoted from Ref. 29. (b): Doping dependence of the $\gamma(0)$ in
(a) normalized by $\gamma_N$, $\gamma(0)/\gamma_N$. The same for
Tl-2201 (diamonds) is quoted from Ref. 36. The normalized residual
spin Knight shift in LSCO,\cite{SO} $N_{\mathrm{res}}/N_N$, is also
shown for comparison (down-triangles). } \label{fig5}
\end{figure}

Now we examine the implication of the residual term $\gamma(0)$ in
zero-field. Figure 5(a) summarizes the doping dependence of
$\gamma(0)$, where the values from previous studies are also
included.\cite{HHW2,MN} For comparison, the normal-state SH
coefficient $\gamma_N$ in the corresponding doping region is shown
together.\cite{NM} We can see that the $\gamma$(0) increases with
doping up to $x=0.259$. For the nonsuperconducting $x=0.290$ sample,
the $\gamma(0)$ is actually the $\gamma_N$, which shows good
consistency with the previous value. Note that for $x=0.259$, the
$\gamma(0)$ is already close to the reported $\gamma_N$. Close to
the optimal doping point the small $\gamma(0)$ may be attributed to
the impurity scattering by which a finite DOS is generated for a
$d$-wave superconductor. However the large $\gamma(0)$ appearing
beyond $x\sim0.22$ cannot be simply attributed to this reason. This
can be understood by having an estimation on the impurity scattering
induced DOS $\gamma_{\mathrm{res}}^{\mathrm{imp}}$ in the
superconducting state, which has the relation
$\gamma_{\mathrm{res}}^{\mathrm{imp}}/\gamma_{N}=(2\gamma_{0}/\pi\Delta_0)\ln(\Delta_0/\gamma_0)$
with $\gamma_0$ the pair breaking parameter.\cite{GP} Also in the
unitarity limit, $\gamma_0\simeq0.6\sqrt{\Gamma\Delta_0}$ with
$\Gamma=1/2\tau_0$ the normal state quasiparticle scattering rate
which can be estimated from the residual resistivity
$\rho_{0}=m^{\ast}/ne^2\tau_0$ and the plasma frequency
$\omega_p=\sqrt{ne^2/\epsilon_0m^{\ast}}$. With $\rho_{0}=26$
$\mu\Omega$cm and $\omega_p\simeq1.2$ eV for $x=0.238$,\cite{MS} one
gets $\hbar\Gamma\simeq2.5$ meV. Assuming $\Delta_0=2.14k_BT_c$, we
obtain $\gamma_{\mathrm{res}}^{\mathrm{imp}}/\gamma_N\simeq0.2$ and
therefore $\gamma_{\mathrm{res}}^{\mathrm{imp}}\simeq2.9$
mJmol$^{-1}$K$^{-2}$ for $x=0.238$, which is far below the
$\gamma(0)$.

\begin{figure}
\includegraphics[width=9cm]{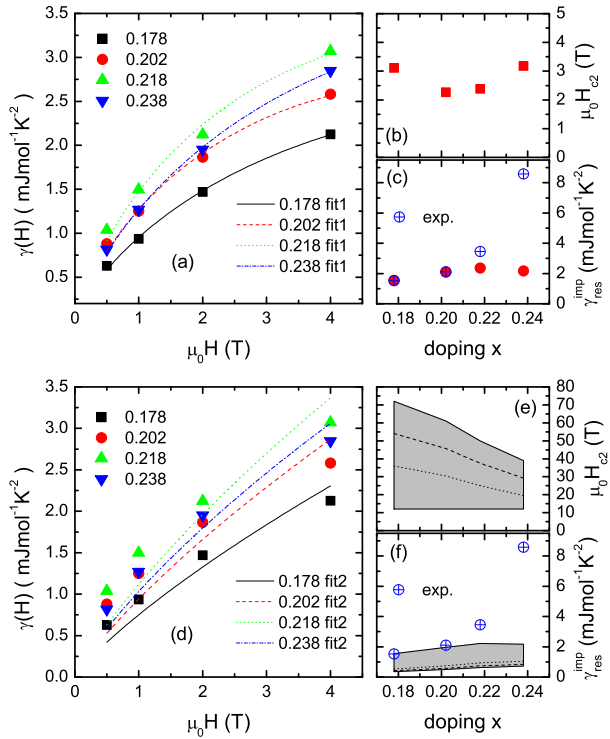}
\caption{(color online) Fit low field $\gamma(H)$ to the $H\ln H$
function (see text). (a): Fit1 with $\Lambda$ and $H_{c2}$ as both
free fitting parameters. The yielded parameters ($\Lambda$
translating to $\gamma_{\mathrm{res}}^{\mathrm{imp}}$) are shown in
(b) and (c). (d): Fit2 with $H_{c2}$ fixed within the values shown
in the shaded region in (e). The dotted and dashed lines in (e)
denote $H_{c2}=1T_c$ and 1.5$T_c$, respectively. The yielded
parameter $\gamma_{\mathrm{res}}^{\mathrm{imp}}$ is shown in (f) as
the shaded region. For comparison, the experimental $\gamma(0)$ is
shown in (c) and (f) (circles).} \label{fig6}
\end{figure}

Furthermore, if the large $\gamma(0)$ is completely induced by the
impurity scattering, the field-induced $\gamma(H)$ at low $H$ is
expected to deviate from the $\sqrt{H}$ dependence and instead show
an $H\ln H$ behavior:
$\gamma(H)\simeq\Lambda(H/H_{c2})\ln[B(H_{c2}/H)]$, where
$\Lambda=\Delta_0a^2\gamma_N/8\gamma_0$ with $B=\pi/2a^2$.\cite{CK}
In Fig. 6 we present the fits to $\gamma(H)$ for $H\leqslant4$ T
with this function. First, we leave $\Lambda$ and $H_{c2}$ as both
free fitting parameters (fit1) and the best fit is shown in Fig.
6(a). As shown in Fig. 6(b), this yields the parameter
$\mu_0H_{c2}<4$ T for all samples $0.178\leqslant x\leqslant0.238$.
The rather low $H_{c2}$ is physically unacceptable. At the same
time, from the parameter $\Lambda$, the coefficient of the residual
specific heat $\gamma_{\mathrm{res}}^{\mathrm{imp}}$ can be
calculated using the expressions and the $\gamma_N$ described above.
It can be seen, for $x\geqslant0.218$, the obtained
$\gamma_{\mathrm{res}}^{\mathrm{imp}}$ is also inconsistent with the
experiment. Secondly, we try to fit the data with $H_{c2}$ fixed
within the values shown in the shaded region in Fig. 6(e) (fit2).
The transport and Nernst effect measurements have indicated that
$\mu_0H_{c2}\sim1.5T_c$ ($H_{c2}$ in Tesla and $T_c$ in Kelvin) for
the overdoped LSCO.\cite{YA,YW06} The current SH suggests
$\mu_0H_{c2}>12$ T for all samples. Hence, in Fig. 6(e) the lower
limit of the shaded region is set to be $\mu_0H_{c2}=12$ T and the
upper limit to be $\mu_0H_{c2}=2T_c$. In this case, we could not
obtain a satisfactory fit to the data, as indicated by the typical
result shown in Fig. 6(d). Again, the obtained
$\gamma_{\mathrm{res}}^{\mathrm{imp}}$ is contradictory to the
measurement (Note, for a given sample, one would obtain the lower
$\gamma_{\mathrm{res}}^{\mathrm{imp}}$ with a higher $H_{c2}$) (Fig.
6(f)). Therefore, it seems that the impurity scattering effect could
not account for the field dependence of the $\gamma(H)$.

The above analysis suggests that in highly doped LSCO the
$\gamma(0)$ mainly comes from contributions other than the impurity
scattering. We attribute it to the presence of nonsuperconducting
metallic regions. This can be corroborated by simultaneously having
a good consistency with the Volovik's relation $\gamma(H)=A\sqrt{H}$
and the very large ratio $\gamma(0)/\gamma_N$ on the single sample
$x=0.238$. Figure 5(b) shows the ratio of $\gamma(0)/\gamma_N$
together with the normalized residual spin Knight shift,\cite{SO}
$N_{\mathrm{res}}/N_N$, another probe of the residual DOS in the
superconducting state. In overdoped Tl2201 the low-temperature SH
has been measured by Loram \emph{et al}. \cite{JWL} and the
$\gamma(\mathrm{5K})/\gamma_N$ is plotted together. We can see that
all these quantities show a rapid increase with overdoping,
indicating a generic property. Actually in LSCO previous results
also showed the rapid increase of $\gamma(0)$ with doping in highly
overdoped region although those experiments were done on
polycrystalline samples.\cite{JWL,NM} One may argue, in LSCO, that
there is a high-temperature tetragonal to orthorhombic structural
transition near $x\simeq0.2$,\cite{HT} which may induce the rapid
increase of $\gamma(0)$. We note that, however, in Pr-doped
LSCO,\cite{WS} this subtle structural transition can be tuned to
much higher doping level, but the superconducting dome remains
unchanged, indicating that the hole concentration rather than the
slight structure distortion plays a dominant role here. Furthermore,
as shown in Fig. 5, a very similar residual $\gamma(0)$ appears in
Tl2201, a system without such a structural transition.

The presence of nonsuperconducting metallic phase implies
immediately a decrease of the superconducting volume fraction. This
can just explain the field dependence of the SH in $x=0.238$ and
0.259 samples. For $x=0.238$, the observed $A$ is even lower than
that for $x=0.218$, implying a significantly reduced superconducting
volume fraction. Taking this into account, the $A$ used to derive
the $\Delta_0$ for this sample, should be corrected roughly as
$A\gamma_N/[\gamma_N-\gamma(0)]$, with the assumption that the
volume ratio is similar to the DOS ratio. The gap value yielded with
this correction is about 3.5 meV, which also scales with $T_c$ in
$d$-wave BCS manner and is plotted in Fig. 4. For $x=0.259$, the
sample shows a large $\gamma(0)$ being close to the $\gamma_N$,
indicating a rather small superconducting volume fraction.

So far, we have shown that in overdoped LSCO, the superconducting
gap decreases with increasing $x$ and at high doping levels, there
exist nonsuperconducting metallic regions at $T\rightarrow0$. Let us
discuss the implications of both effects. Previously it was
suggested that the suppression of $T_c$ in overdoped regime may come
from the increasing pair breaking effect. Our present result,
however, does not support this proposal since $T_c$ is found to
scale with $\Delta_0$ in good agreement with $d$-wave BCS theory,
which implies that the decrease in $T_c$ should originate from an
underlying reduction in the pairing strength. This point may also be
helpful to elucidate the origin of the presence of
nonsuperconducting metallic regions in the overdoped sample, which
is yet unclear. Currently several scenarios have been proposed to
account for this anomalous phenomenon. One is that the overdoped
cuprate may spontaneously phase separate into the hole-poor
superconducting region and the hole-rich normal Fermi liquid region
due to the competition in energy between these two
phases.\cite{YJU2} Another scenario is associated with the
microscopic hole doping state.\cite{YT} It was speculated that in
the overdoped regime the holes were doped directly into the Cu3d
orbital rather O2p,\cite{SW2} which is expected to produce free Cu
spins and/or disturb the antiferromagnetic correlation between Cu
spins. Around these holes, the superconductivity is destroyed,
forming the normal state region. Our present result seems supporting
this scenario with the assumption that the superconductivity is
magnetic in origin and the suppression of $\Delta_0$ originates from
the disturbing of the antiferromagnetic correlation with overdoping.

\section{SUMMARY}

In summary, low-temperature SH in overdoped LSCO single crystals has
revealed two interesting findings: (1) The field-induced SH follows
the prediction of $d$-wave symmetry yielding a gap value $\Delta_0$
approaching closely onto the weak-coupling $d$-wave BCS relation
$\Delta_0=2.14k_BT_c$; (2) At high doping levels, the residual SH
term $\gamma(0)$ rises dramatically with doping, which suggests the
existence of unpaired electrons possible in association with the
normal metallic regions. These discoveries may carry out a common
feature in cuprate superconductors and give important clues to the
high-$T_c$ pairing mechanism.

\begin{acknowledgments}
This work is supported by the National Science Foundation of China,
the Ministry of Science and Technology of China (973 project No:
2006CB601000, 2006CB921802), and Chinese Academy of Sciences
(Project ITSNEM).
\end{acknowledgments}

\end{document}